\documentclass[aps,prd,amsmath,nofootinbib,preprintnumbers,superscriptaddress]{revtex4-1}
\usepackage{epsfig,slashed,nicefrac}
\usepackage[utf8]{inputenc}
\usepackage{color}
\usepackage{float} 
\usepackage{subfig} 
\usepackage{eqnarray,amsmath}

\newcommand{\ho}{{\sc HELAC-Onia}}

\newcommand{\FeynArts}{{\sc FeynArts}}
\newcommand{\FeynCalc}{{\sc FeynCalc}}
\newcommand{\Mathematica}{{\sc Mathematica}}
\newcommand{\LoopTools}{{\sc LoopTools}}
\newcommand{\QCDLoop}{{\sc QCDLoop}}

\def\be{\begin{equation*}}
\def\ee{\end{equation*}}
\def\bsp#1\esp{\begin{split}#1\end{split}} 
\def\bpm{\begin{pmatrix}}
\def\epm{\end{pmatrix}}

\begin{document}

\preprint{CERN-TH-2016-096}

\title{Complete Study of Hadroproduction of a $\Upsilon$ Meson Associated with a Prompt $J/\psi$}

\author{Hua-Sheng Shao}
\affiliation{Theoretical Physics Department, CERN, CH-1211 Geneva 23, Switzerland}

\author{Yu-Jie Zhang}
\affiliation{Beijing Key Laboratory of Advanced Nuclear Energy Materials and Physics, and School of Physics, Beihang University, Beijing 100191, China}
\affiliation{CAS Center for Excellence in Particle Physics, Beijing 100049, China}

\date{\today}

\begin{abstract}
We present the first complete study of $\Upsilon$ and prompt $J/\psi$ production from single-parton scattering, including the complete $\mathcal{O}(\alpha_S^6)$ color-singlet contribution, the $\mathcal{O}(\alpha_S^2\alpha^2)$ electroweak contribution, the complete nonrelativistic S-wave and P-wave color-octet contribution as well as the feeddown contribution. Our study was motivated by the recent evidence reported by D0 Collaboration of prompt $J/\psi$ and $\Upsilon$ simultaneous production at the Tevatron. With our complete evaluation, we are able to refine the determination of the double parton scattering contribution made by D0 Collaboration. We find that the effective cross section characterizing the importance of double-parton scatterings is $\sigma_{\rm eff}\le 8.2$ mb at $68\%$ confidence level from the D0 measurement.
\end{abstract}

\maketitle

\textit{Introduction} -- Heavy quarkonium production at colliders have been extensively studied (see e.g. the reviews Refs.~\cite{Brambilla:2010cs,Andronic:2015wma}).  On one hand, it contains a rich physics which is interesting on its own; on the other hand, quarkonia are used as tools of many facets of the standard model. Despite the lack of consensus about its dominant mechanism, associated-quarkonium production has attracted considerable theoretical attention because it provides a good opportunity to study the multiple parton interactions. Indeed, associated-quarkonium production offers relatively large yields and is usually experimentally clean to measure. The understanding of multiple parton interactions in hadron-hadron collisions is very important, for it can be an important background of multiparticle final-state processes. It impacts both the measurements of standard model particles and the searching for new physics signals. Many quarkonium-associated production processes, such as $J/\psi+W$~\cite{Aad:2014rua}, $J/\psi+Z$~\cite{Aad:2014kba}, $J/\psi+charm$~\cite{Aaij:2012dz}, $\Upsilon+charm$~\cite{Aaij:2015wpa}, and $J/\psi+J/\psi$~\cite{Abazov:2014qba,Khachatryan:2014iia,Lansberg:2014swa}, seem to be dominated by double-parton scattering (DPS). However, one should keep in mind that before concluding for DPS dominance, one should always carefully examine the single-parton scattering (SPS) contributions. The situation is usually unclear since the quantification of the SPS for quarkonium-associated production is often  challenging.

In this Letter, we focus on the theoretical studies of prompt $\psi+\Upsilon$ production motivated by the recent claim made by the D0 Collaboration~\cite{Abazov:2015fbl} of the first evidence of $J/\psi+\Upsilon(1S,2S,3S)$ production at hadron colliders. Unlike $J/\psi$-pair or $\Upsilon$-pair production ~\cite{Kartvelishvili:1984ur,Humpert:1983yj,Vogt:1995tf,Li:2009ug,Qiao:2009kg,Ko:2010xy,Berezhnoy:2011xy,Li:2013csa,Lansberg:2013qka,Sun:2014gca,Lansberg:2014swa}, neither $\mathcal{O}(\alpha_S^4)$ nor $\mathcal{O}(\alpha_S^5)$ contributions survive in color-singlet model (CSM). The process is thus sometimes considered as a golden observable to probe the so-called color-octet mechanism (COM)~\cite{Ko:2010xy}, which can be seen as a relativistic correction via high Fock state contribution in the meson wave function. However, the color-octet (CO) contributions were quite underestimate until predictions were made for AFTER@LHC energies $\sqrt{s}=115$ GeV~\cite{Lansberg:2015lva} relying on the automation in \ho~\cite{Shao:2012iz,Shao:2015vga}. The approximated loop-induced (LI) contribution in CSM at $\mathcal{O}(\alpha_S^6)$ was estimated in Ref.~\cite{Likhoded:2015zna} within the specific limit $\hat{s}\gg |\hat{t}| \gg m^2_{\psi,\Upsilon}$, where $\hat{s}$ and $\hat{t}$ are the Mandelstam variables. However, the exact calculations of the complete SPS contributions were absent in the literature.  

The aim of this Letter is to present the first complete study of the simultaneous production of prompt $\psi$~\footnote{$\psi$ production from b-hadron decay is excluded.} and $\Upsilon$ mesons by including all leading contributions, at order $\mathcal{O}(\alpha_S^6)$  or equivalent.

\medskip

\textit{Theoretical framework} --In general, the SPS cross section for the simultaneous production of charmonium $\mathcal{C}$ and bottomonium $\mathcal{B}$ in the nonrelativistic limit can be written as
\be\bsp
& \sigma(h_1h_2\rightarrow \mathcal{C}+\mathcal{B}+X) = \sum_{a,b,n_1,n_2}f_{a/h_1}\otimes f_{b/h_2} \\
& \otimes \hat{\sigma}(ab\rightarrow c\bar{c}[n_1]+b\bar{b}[n_2]+X)\langle O^{\mathcal{C}}(n_1) \rangle \langle O^{\mathcal{B}}(n_2) \rangle,
\esp \ee
where $f_{a/h}$ is the parton distribution function (PDF), $\langle O^{\mathcal{Q}}(n) \rangle$ is the nonperturbative long-distance matrix element (LDME) of the quarkonium $\mathcal{Q}$,\footnote{It has the simple physical probability interpretation at leading order.} and $\hat{\sigma}(ab\rightarrow c\bar{c}[n_1]+b\bar{b}[n_2]+X)$ is the perturbatively calculable short-distance coefficient for the simultaneous production of the charm-quark pair in the Fock state $n_1$ and the bottom-quark pair in the Fock state $n_2$. The contributions from various Fock states can be organized in the nonrelativistic limit; i.e., the importance can be ordered in powers of $v_q$, where $v_q$ is the relative velocity of the heavy-flavor quark pair $q\bar{q}$  that formed the heavy quarkonium $\mathcal{Q}$. Approximately, one has $v_c^2\simeq 0.3$ in charmonium and $v_b^2\simeq 0.1$ in bottomonium. The leading contribution in $v_q$ for S-wave quarkonium is from the color-singlet (CS) production. 

The $\mathcal{O}(\alpha_S^4)$ and  $\mathcal{O}(\alpha_S^5)$ contributions to $\Upsilon$ and $\psi$ direct production in CSM vanish because of P-parity and C-parity conservation. Other production mechanisms {\it a priori} considered to be subleading can be relevant. In the following, we will consider all the contributions which can compete with the $\mathcal{O}(\alpha_S^6)$ CSM ones. Besides the possible DPS contribution, there are five relevant classes of production mechanisms, which are summarized in Table.\ref{tab:spsmech}.

\begin{figure}[hbt!]
\centering
\subfloat[DR]{\includegraphics[width=.2\columnwidth,draft=false]{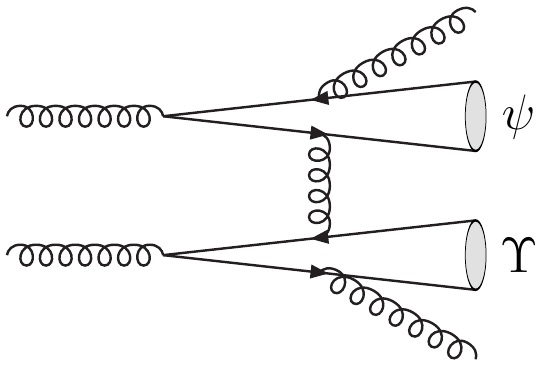}\label{diagram-a}}
\subfloat[LI]{\includegraphics[width=.2\columnwidth,draft=false]{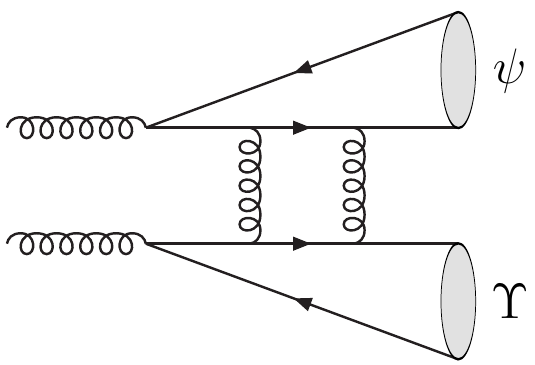}\label{diagram-b}}
\subfloat[EW]{\includegraphics[width=.2\columnwidth,draft=false]{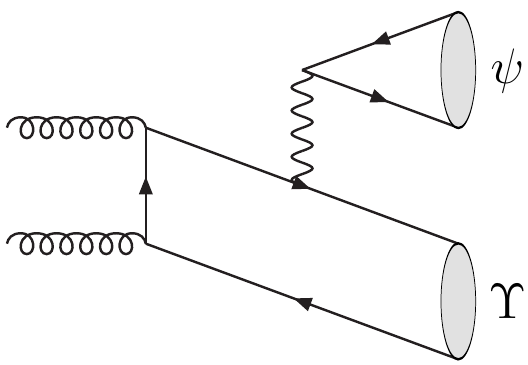}\label{diagram-c}}
\subfloat[COM]{\includegraphics[width=.2\columnwidth,draft=false]{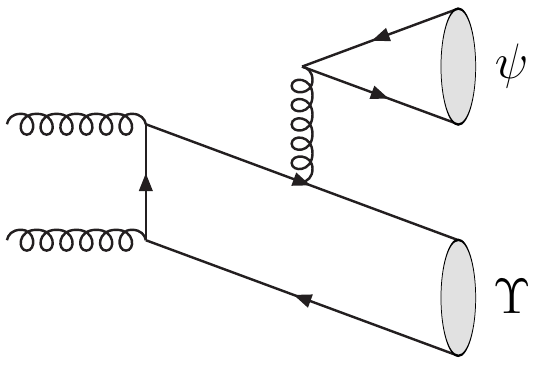}\label{diagram-d}}
\subfloat[FD]{\includegraphics[width=.2\columnwidth,draft=false]{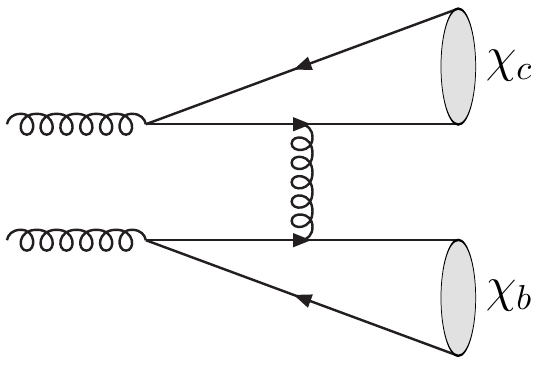}\label{diagram-e}}
\caption{Diagrams for the hadroproduction of prompt $\psi+\Upsilon$ via SPS at $\mathcal{O}(\alpha_S^6)$ (a-b), at $\mathcal{O}(\alpha_S^2\alpha^2)$ (c) and 
 $\mathcal{O}(\alpha_S^4 v_c^iv_b^j,i+j\ge 4)$ (d-e).}
\label{diagrams} \vspace*{-0.5cm}
\end{figure}

As announced, we have evaluated the complete $\mathcal{O}(\alpha_S^6)$ CS contribution, which includes the double real (DR) emission diagrams $gg\rightarrow \psi+\Upsilon+gg$ and the LI diagrams $gg\rightarrow \psi+\Upsilon$ ( see respectively, Figs.~\ref{diagram-a} and \ref{diagram-b} for representative graphs). Such a complete calculation has never been performed before, in particular, as what regards the LI contributions, which are beyond tree-level techniques. For the latter, we have performed two independent computations. For the LI contribution: we have used \FeynArts~\cite{Hahn:2000kx} to generate the one-loop amplitude and applied two methods to calculate it.  The first consisted in using \FeynCalc~\cite{Mertig:1990an,Shtabovenko:2016sxi}  and \LoopTools~\cite{Hahn:1998yk} to calculate the loop integrals in mass regularization, the second in using the in-house \Mathematica\ program to reduce the dimensional-regularized one-loop  tensor integrals and to evaluate the one-loop scalar integrals with the help of \QCDLoop~\cite{Ellis:2007qk}. 

\begin{table*}[!t]
\renewcommand{\arraystretch}{1.4}
\setlength{\tabcolsep}{12pt}
 \begin{tabular}{c||ccc}
    Label & \ho\ 2.0 syntax & First order  & Description\\
    \hline \hline
   DR & {\tt g g > cc$\sim$(3S11) bb$\sim$(3S11) g g}  &
         $\mathcal{O}(\alpha_S^6)$ & Double Real (DR) CS contribution\\
   LI & {\tt addon 8}  &
         $\mathcal{O}(\alpha_S^6)$ & Loop-Induced (LI) CS contribution\\
   EW & {\tt p p > cc$\sim$(3S11) bb$\sim$(3S11)} & $\mathcal{O}(\alpha_S^2\alpha^2)$ & ElectroWeak (EW) CS contribution\\
   INTER & {\tt addon 8}  & $\mathcal{O}(\alpha_S^4\alpha)$ &INTERference (INTER) between LI and EW\\
   COM & {\tt g g > jpsi y(1s)} & $\mathcal{O}(\alpha_S^4v_c^iv_b^j),i+j\ge 4$ & CO $\mathcal{O}(\alpha_S^4)$ contribution
\end{tabular}
\renewcommand{\arraystretch}{1.0}
\caption{\small \label{tab:spsmech}Various SPS hadroproduction channels of $\psi+\Upsilon$ we considered, where we also present the correspoding syntax to compute with \ho\ 2.0.}
\end{table*}

As what regards the electroweak (EW) contribution in the CSM, they appear at $\mathcal{O}(\alpha_S^2\alpha^2)$, where $\alpha$ is the electromagnetic coupling constant. If one considers that $\alpha\sim \alpha_S^2$, $\mathcal{O}(\alpha_S^2\alpha^2)$ contributions are on the same order as $\mathcal{O}(\alpha_S^6)$ CS contributions and should thus be taken into account in our calculations. We also naturally consider the interference (INTER) term between  LI diagrams and EW diagrams. Such an $\mathcal{O}(\alpha_S^4\alpha)$ interference term is also on the same order as the $\mathcal{O}(\alpha_S^6)$ CS term considering $\alpha\sim \alpha_S^2$. 

The last possible relevant contribution from SPSs is from the CO contributions. They consist of both the CS+CO and CO+CO channels with 66 different nonvanishing channels (see the Appendix of Ref.~\cite{Lansberg:2015lva}). We have taken into account all S-wave and P-wave Fock states up to $\mathcal{O}(v_q^4)$ in nonrelativistic QCD~\cite{Bodwin:1994jh}, including the $\chi_c+\chi_b$ production (diagram like Fig.\ref{diagram-e}),~\footnote{The simultaneous production of $\chi_c$ and $\chi_b$ mesons was first considered in Ref.~\cite{Likhoded:2015zna}.} which contribute to the feeddown (FD) yield. The calculation of this piece will be restricted to $\mathcal{O}(\alpha_S^4)$. We labeled it as ``COM". Such a complete computation of the CO contributions was first presented in Ref.~\cite{Lansberg:2015lva} for the kinematics of AFTER@LHC~\cite{Brodsky:2012vg}. If one assumes $v_q^2\sim \alpha_S$, we arrived at the same order $\mathcal{O}(\alpha_S^6)$ again.

Besides the SPSs, one can expect a significant amount of the yields from the DPS, in which $\psi$ and $\Upsilon$ are produced separately in two partonic scattering processes. Despite the absence of a complete proof of DPS factorization,~\footnote{The cancellation of the Glauber gluon in DPS has been proven recently in the double Drell-Yan process~\cite{Diehl:2015bca}.} we use as a first approximation the well-known ``pocket formula", which assumes the complete factorization and independence of both hard processes
{\small
\begin{equation}\bsp
\hspace{-0.5cm}
& \sigma^{\rm DPS}(h_1h_2\rightarrow \mathcal{C}+\mathcal{B}+X) =\frac{\sigma(h_1h_2\rightarrow \mathcal{C}+X)\sigma(h_1h_2\rightarrow \mathcal{B}+X)}{\sigma_{\rm eff}},\label{eq:dps}
\esp\end{equation}
}
where the effective parameter $\sigma_{\rm eff}$ accounts for an effective transverse overlapping area of the two initial colliding hadrons $h_1$ and $h_2$. Such an approximation is considered to be reasonable in the low-x regime, i.e., at LHC energies. DPS studies are still currently in an ``exploration" phase, and DPS factorization still needs to be checked case by case. In Ref.~\cite{Abazov:2015fbl}, the D0 Collaboration extracted $\sigma_{\rm eff}=2.2\pm 0.7 (stat) \pm 0.9 (syst)$ mb from their $J/\psi+\Upsilon$ measurement by assuming a negligible SPS contribution. It can be conservatively thought as the lower-limit value of $\sigma_{\rm eff}$ if one assumes the universality of $\sigma_{\rm eff}$. Along the lines of Refs.~\cite{Kom:2011bd,Lansberg:2014swa,Lansberg:2015lva}, we use a data-driven procedure as input for the single quarkonium production cross sections.

\medskip

\textit{Results} --
All of the contributions described above have been implemented in the framework of \ho~\cite{Shao:2012iz,Shao:2015vga}, with the syntax listed in Table.\ref{tab:spsmech}. For the numerical evaluations, we have taken the charm (bottom) quark mass to be half of the mass of the $J/\psi$ ($\Upsilon(1S)$) meson $m_c=m_{J/\psi}/2\simeq 1.55$ GeV ($m_b=m_{\Upsilon(1S)}/2\simeq 4.73$ GeV). The renormalization scale $\mu_R$ and the factorization scale $\mu_F$ dependence has been estimated by the independent variations $\mu_0/2<\mu_R,\mu_F<2\mu_0$, where the central scale $\mu_0$ is taken to be the half of the summed transverse masses of the charmonium and the bottomonium, i.e.,
$\mu_0=\left(\sqrt{(p_T^{\mathcal{C}})^2+4m_c^2}+\sqrt{(p_T^{\mathcal{B}})^2+4m_b^2}\right)/2$. We have used the next-to-leading order (NLO) PDF4LHC15 set~\cite{Butterworth:2015oua,Dulat:2015mca,Harland-Lang:2014zoa,Ball:2014uwa,Gao:2013bia,Carrazza:2015aoa,Watt:2012tq} available in LHAPDF6.1.6~\cite{Buckley:2014ana} as our default PDF. We also estimate the PDF uncertainty by considering its 30 error eigenvectors. We have found that the main theoretical uncertainties in the SPS contributions are the renormalization scale $\mu_R$ and factorization scale $\mu_F$ dependence. 

The default values for the CS LDMEs we used were estimated in Ref.~\cite{Eichten:1995ch} with the Buchmuller-Tye potential~\cite{Buchmuller:1980su}. In order to include the feeddown contributions from the higher-excited quarkonium-state decay, we take the world-averaged branching ratios from the PDG~\cite{Agashe:2014kda}. To fully take into account the feeddown contributions, multiple transitions, in particular, for the bottomonium, need to be considered. This amounts to tedious computations. To do so, we have implemented a general algorithm to compute the feeddown contributions from the multiple radiative transitions in  multiple-onium production processes in \ho\ 2.0~\cite{Shao:2015vga}.

On the other hand, the estimation of DPS via the ``pocket formula" Eq.(\ref{eq:dps}) requires the knowledge of the single quarkonium production cross sections. They are estimated by fitting the crystal ball function~\cite{Kom:2011bd} to the Tevatron data with the MSTW2008NLO set~\cite{Martin:2009iq} for the Tevatron production processes and to the LHC data with the same PDF for the LHC production processes. For more details, we guide the reader to Refs.~\cite{Lansberg:2014swa,Lansberg:2015lva}. In the present case, we have found that the template dependence is mild. Indeed, the inclusion of LHC data (vs Tevatron only) to determine the single quarkonium inputs alters the DPS cross section at the Tevatron by $12\%$.


The D0 fiducial cross section is defined by requiring the muons from $J/\psi$ and $\Upsilon$ decay to have at least $p_T(\mu^{\pm})>2$ GeV and $|\eta(\mu^{\pm})|<2.0$~\cite{Abazov:2015fbl}. The full spin correlations in $\psi(\Upsilon)\rightarrow \mu^+\mu^-$ and in $\chi_{c} (\chi_b)\rightarrow \psi (\Upsilon)+\gamma$~\cite{Shao:2012fs,Shao:2014yfa} have been used in order to get the correct muon acceptance. Since the $p_T$ cut on the muon is not very high, one might doubt that the initial $k_T$ smearing will marginally change the acceptance. We have checked that the fraction of events passing the fiducial cuts is affected by the initial $k_T$ smearing at most at $20\%$ level when $\langle k_T \rangle\le 3$ GeV.

The SPS cross sections for direct $\psi(1S,2S)+\Upsilon(1S,2S,3S)$ production in the D0 fiducial region are presented in Table.\ref{tab:xsdirect}, where we detailed the contributions from five different production mechanisms. In $\sigma_{\rm EW}$, we have included both gluon-gluon initial state and quark-antiquark initial state, while only gluon-gluon initial state EW diagrams can interfere with the loop diagrams in $\sigma_{\rm INTER}$ and the contribution from the quark-antiquark initial state is quite small because of $\alpha$ and PDF suppressions.  We want to point out that the negative contribution from $\sigma_{\rm INTER}$ almost cancels with the positive contribution from $\sigma_{\rm EW}$ in the D0 fiducial region. However, such a cancellation does not happen in the LHCb acceptance (i.e., $2.0<y_{J/\psi,\Upsilon}<4.5$ at $\sqrt{s}=13$ TeV proton-proton collision). $\sigma_{\rm DR}$ is small because the extra radiations of hard gluons tend to increase the invariant mass of the produced system and thus a larger $x$ in PDF. $\sigma_{\rm COM}$ and $\sigma_{\rm LI}$ tend to be the largest SPS $\psi$ and $\Upsilon$ contributions. The renormalization and factorization scale dependence is the main theoretical uncertainty, which can be reduced by including higher-order QCD corrections.

\begin{table}
\begin{footnotesize}
\hspace{-0.5cm}
\begin{tabular}{c|c|c|c}
\hline
\rule{0pt}{3ex}
& & 
 $J/\psi$ & $\psi(2S)$\\[1mm] 
\hline\hline\rule{0pt}{3ex}
& $\Upsilon(1S)$  
  & $3.58^{+233\%}_{-66.4\%}\pm4.4\%$ 
  & $2.34^{+233\%}_{-66.4\%}\pm4.4\%$\\ 
DR & $\Upsilon(2S)$  
  & $1.78^{+233\%}_{-66.4\%}\pm4.4\%$
  & $1.17^{+233\%}_{-66.4\%}\pm4.4\%$\\ 
 & $\Upsilon(3S)$ 
  & $1.36^{+233\%}_{-66.4\%}\pm4.4\%$
  & $0.894^{+233\%}_{-66.4\%}\pm4.4\%$\\ 
\hline\rule{0pt}{3ex}
& $\Upsilon(1S)$  
  & $56.2^{+264\%}_{-70.2\%}\pm4.7\%$ 
  & $36.8^{+264\%}_{-70.2\%}\pm4.7\%$\\ 
LI & $\Upsilon(2S)$  
  & $28.0^{+264\%}_{-70.2\%}\pm4.7\%$ 
  & $18.4^{+264\%}_{-70.2\%}\pm4.7\%$\\ 
 & $\Upsilon(3S)$ 
  & $21.4^{+264\%}_{-70.2\%}\pm4.7\%$ 
  & $14.0^{+264\%}_{-70.2\%}\pm4.7\%$\\ 
\hline
\rule{0pt}{3ex}
& $\Upsilon(1S)$  
  & $15.8^{+75.4\%}_{-46.4\%}\pm4.6\%$ 
  & $10.4^{+75.4\%}_{-46.4\%}\pm4.6\%$ \\ 
EW & $\Upsilon(2S)$  
  & $7.90^{+75.4\%}_{-46.4\%}\pm4.6\%$
  & $5.18^{+75.4\%}_{-46.4\%}\pm4.6\%$\\ 
 & $\Upsilon(3S)$ 
  & $6.04^{+75.4\%}_{-46.4\%}\pm4.6\%$ 
  & $3.96^{+75.4\%}_{-46.4\%}\pm4.6\%$\\ 
\hline
\rule{0pt}{3ex}
& $\Upsilon(1S)$  
  & $-16.6^{+162\%}_{-62.0\%}\pm 4.8\%$ 
  & $-10.9^{+162\%}_{-62.0\%}\pm 4.8\%$ \\ 
INTER & $\Upsilon(2S)$  
  & $-8.29^{+162\%}_{-62.0\%}\pm 4.8\%$
  & $-5.43^{+162\%}_{-62.0\%}\pm 4.8\%$\\ 
 & $\Upsilon(3S)$ 
  & $-6.34^{+162\%}_{-62.0\%}\pm 4.8\%$ 
  & $-4.15^{+162\%}_{-62.0\%}\pm 4.8\%$\\ 
\hline
\rule{0pt}{3ex}
& $\Upsilon(1S)$  
  & $409^{+138\%}_{-56.7\%}\pm4.4\%$ 
  & $174^{+138\%}_{-56.8\%}\pm4.4\%$\\ 
COM & $\Upsilon(2S)$  
  & $135^{+139\%}_{-57.0\%}\pm4.4\%$ 
  & $57.6^{+139\%}_{-57.1\%}\pm4.4\%$\\ 
 & $\Upsilon(3S)$ 
  & $197^{+137\%}_{-56.6\%}\pm4.4\%$ 
  & $84.1^{+138\%}_{-56.7\%}\pm4.4\%$\\  
\hline
\end{tabular}
\caption{Cross sections (in units of fb) of direct $\psi+\Upsilon$ production via SPS with $\sqrt{s}=1.96$ TeV proton-antiproton collisions in the D0 fiducial region. The quoted errors are from renormalization or factorization scale dependence and PDF uncertainty, respectively.}
\label{tab:xsdirect} 
\end{footnotesize}
\end{table}

After including the feeddown contribution, which can be as large as the direct production contribution, the cross sections for the prompt $J/\psi$ and $\Upsilon$ production in the fiducial regions of the D0 and LHCb ongoing measurements~\cite{LHCb:2016xx} are given in Table.\ref{tab:xsprompt}. The dependence of $\sigma_{\rm COM}$ on LDMEs is estimated by considering four different sets of LDMEs for charmonium and bottomonium production, which can be described as follows:
\begin{itemize}
\item{Set I:} This set is based on the LDMEs presented in Ref.~\cite{Kramer:2001hh}, where the LDMEs of $\chi_b(3P)$ are set to 0. This set is the default one in our discussion.
\item{Set II:} This set is based on the LDMEs of charmonium and bottomonium extracted in Ref.~\cite{Sharma:2012dy}, where the contribution of $\chi_b(3P)$ was also ignored.
\item{Set III:} Unlike the previous two sets, we used the LDMEs extracted from NLO analyses. The LDMEs of $J/\psi=\psi(1S),\psi(2S),\chi_c$ are taken from Ref.~\cite{Shao:2014yta}, while those of $\Upsilon(nS),\chi_b(nP),1\le n\le 3$ are from Ref.~\cite{Han:2014kxa}. 
\item{Set IV:} A second set of LDMEs based on NLO calculations is taken from Ref.~\cite{Gong:2012ug} for charmonia and from Ref.~\cite{Feng:2015wka} for bottomonia.~\footnote{We do not use other fits~\cite{Butenschoen:2011yh,Bodwin:2014gia} based on NLO calculations because they do not include the feeddown contribution and no corresponding LDMEs for bottomonia are available.}
\end{itemize}
The cross sections $\sigma(J/\psi+\Upsilon(1S,2S,3S)){\rm Br}(J/\psi\rightarrow \mu^+\mu^-){\rm Br}(\Upsilon\rightarrow \mu^+\mu^-)$ with the above four sets are presented in Table.\ref{tab:xsprompt}. Although the dependence of LDMEs is quite big, we can take the value with set I as the conservative upper-limit contribution from this piece, which is, however, still a factor of 10 smaller than the measurement by the D0 Collaboration. One interesting feature we have noticed is that due to the presence of the negative LDMEs, set II and set IV will result in negative cross sections, for example, in the direct production of $J/\psi+\Upsilon(2S)$. Negative cross sections, as opposed to negative LDMEs, are unphysical. Yet, this is reminiscent of the conclusion of Ref.~\cite{Li:2014ava} that associated production channels introduce new constraints on the CO LDMEs and, in fact, nearly rule out fits yielding to negative LDMEs values.

\begin{figure}
\vspace{-1cm}
\hspace{-2cm}
  \includegraphics[width=0.8\columnwidth]{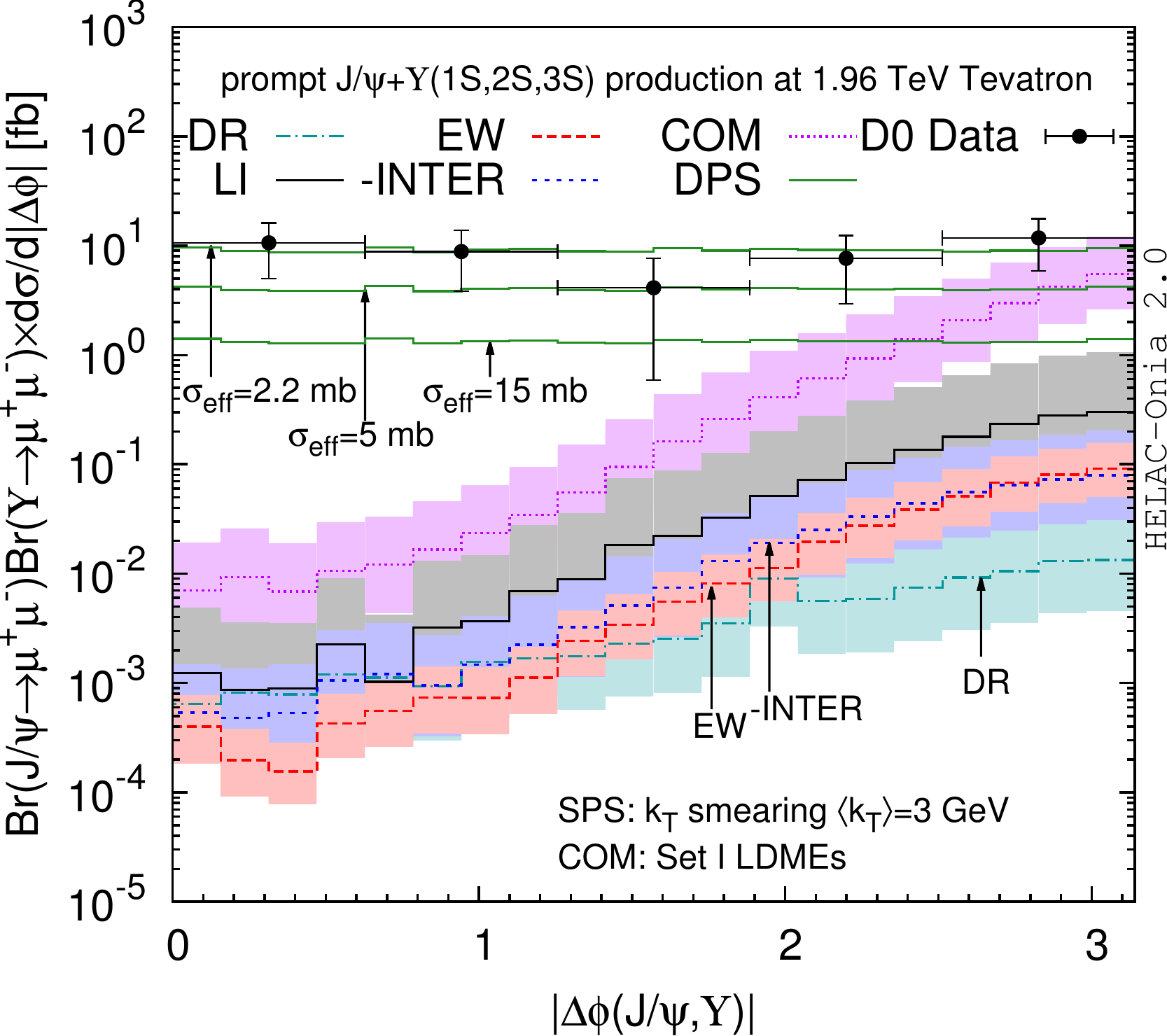}\\
  \hspace{-2cm}
\includegraphics[width=0.8\columnwidth]{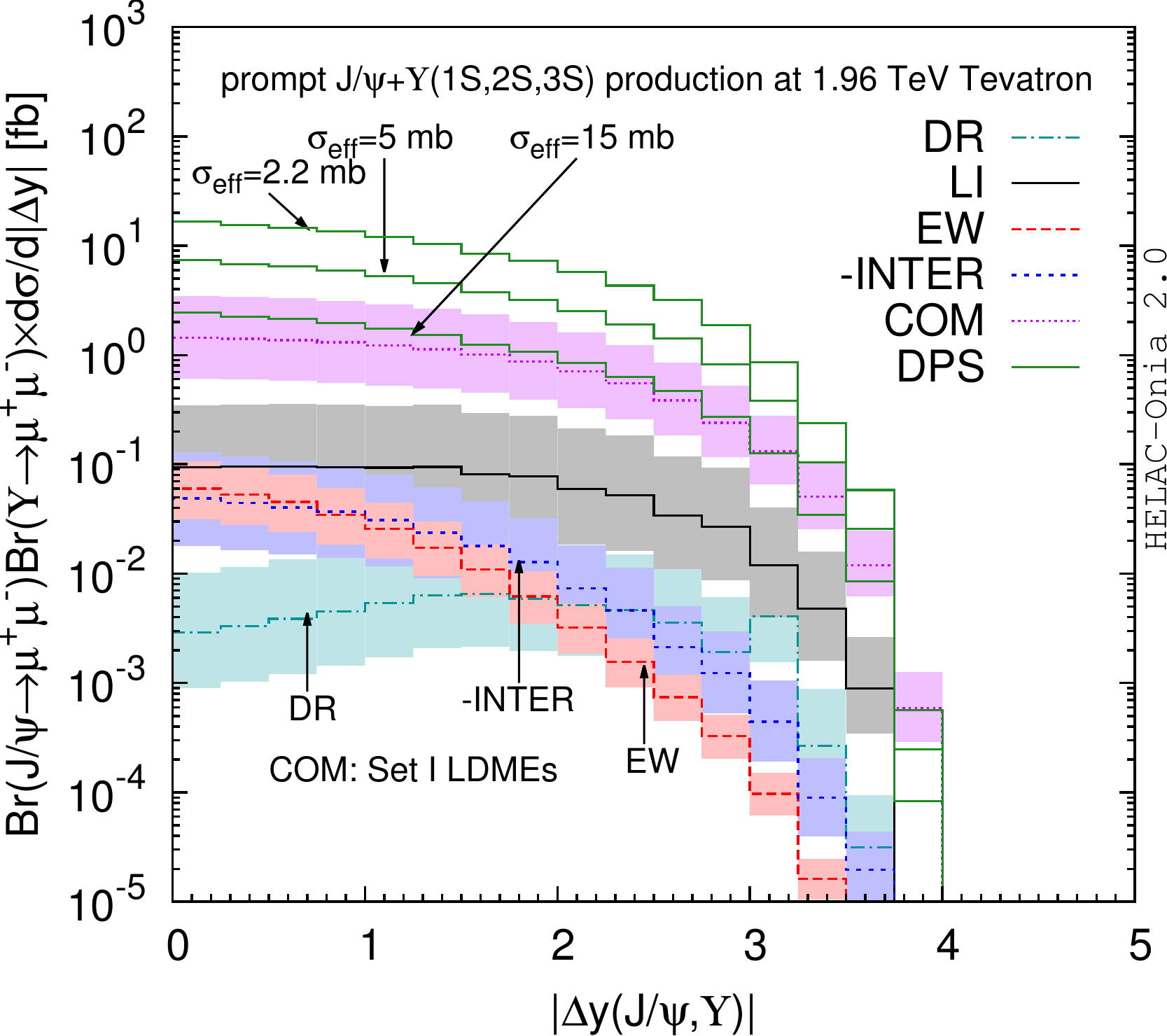}
  \caption{\small \label{fig:dist}Differential distributions of prompt $J/\psi+\Upsilon$ production at the Tevatron within the D0 fiducial region.}
\end{figure}

Two differential distributions for prompt $J/\psi$ and $\Upsilon$ production at the Tevatron are presented in Fig.\ref{fig:dist}, while the other differential distributions as well as the predictions for LHCb are given as Supplementary Material. Because of the limited statistics, the D0 Collaboration only reported on the distribution of the azimuthal angle $\Delta \phi(J/\psi,\Upsilon)$ between $J/\psi$ and $\Upsilon$. The uncorrelation between $J/\psi$ and $\Upsilon$ in DPS proudction results in the flat distribution. On the contrary, the correlations from SPSs are expected to generate a nontrivial distribution. However, it was already pointed out~\cite{Kom:2011bd} that such a distribution may significantly depend on the initial or intrinsic $k_T$ smearing. We use a Gaussian distribution with $\langle k_T\rangle=3$ GeV to mimic the smearing effect. The bins of $\Delta \phi(J/\psi,\Upsilon)\rightarrow 0$ are populated after including such a smearing, especially for the events from $2\rightarrow 2$ topologies, which usually are at $\Delta \phi(J/\psi,\Upsilon)=\pi$. Similar to the total cross section, the differential SPS cross sections alone are not enough to account for the experimental data, which clearly means the necessity of the DPS contribution. Three DPS curves with $\sigma_{\rm eff}=2.2,5,15$ mb are given. Unlike the double $J/\psi$ production, the shape of the distribution in rapidity difference between $J/\psi$ and $\Upsilon$ is no longer a good discrimination between DPS and SPSs; it can be clearly seen from the lower panel of Fig.\ref{fig:dist} that the COM, LI, and DR curves are as broad as the DPS curves. Its physical explaination can be attributed to the contributions from the $t-$channel gluon exchanging diagrams. In the differential distributions, the contributions from INTER and EW are almost canceled again in the D0 fiducial region. In Fig.\ref{fig:dist}, we have multiplied a factor of $-1$ to the INTER curves to make them visible.

Having now at hand a reliable estimation of the SPS yields, we can refine the extraction of $\sigma_{\rm eff}$ related to the DPS yield needed to reproduced the data. The lower-limit value of $\sigma_{\rm eff}$ in $J/\psi+\Upsilon$ was estimated by the D0 Collaboration to be $2.2\pm1.14$ mb in Ref.~\cite{Abazov:2015fbl}, by assuming no SPS contributions at all. Its upper-limit value at $68\%$ confidence level can be extracted by saturating the DPS to $(\sigma^{\rm D0}-\delta \sigma^{\rm D0})-(\sigma^{\rm SPS}+\delta \sigma^{\rm SPS})$, where $\sigma^{\rm D0}$($\sigma^{\rm SPS}$) are the central values of the D0 measured cross section (the calculated SPS cross sections) and $\delta \sigma$ are the corresponding 1 standard deviations. Doing so, we get $\sigma_{\rm eff}\le 8.2$ mb at $68\%$ confidence level. Such a value is very close to the central value needed to explain the excess of the CMS $J/\psi+J/\psi$ data at large rapidity difference $|\Delta y(J/\psi,J/\psi)|$ over the SPS yield~\cite{Lansberg:2014swa} and confirms the DPS contributions are a key element in explaining quarkonium-associated production at the LHC.


\begin{table*}
\begin{normalsize}
\hspace{0cm}
\begin{tabular}{c|c|c|c|c|c|c|c|c}
\hline\rule{0pt}{3ex}
  Experiment & \multicolumn{4}{c|}{CSM} & \multicolumn{4}{c}{COM}\\[1mm]
  & DR & LI & EW & INTER &
 Set I & Set II & Set III & Set IV \\[1mm] 
\hline\hline\rule{0pt}{3ex}
 D0: $27\pm42.2\%$  & $0.0146^{+233\%}_{-66.6\%}$ &$0.229^{+264\%}_{-70.4\%}$
  & $0.065^{+75.5\%}_{-46.6\%}$ & $-0.068^{+162\%}_{-62.2\%}$ & $2.96^{+135\%}_{-56.2\%}$
  & $1.41^{+160\%}_{-77.6\%}$
  & $1.80^{+143\%}_{-58.0\%}$ 
 & $0.418^{+144\%}_{-58.3\%}$\\
\hline
\rule{0pt}{3ex}
 LHCb  & $0.255^{+391\%}_{-79.7\%}$ & $6.05^{+436\%}_{-82.2\%}$
  & $1.71^{+135\%}_{-65.2\%}$ & $-3.23^{+262\%}_{-75.9\%}$ & $38.8^{+238\%}_{-73.0\%}$
  & $21.2^{+243\%}_{-73.6\%}$
  & $28.1^{+243\%}_{-73.8\%}$
 & $6.57^{+243\%}_{-73.9\%}$\\
\hline
\end{tabular}
\caption{Cross sections $\sigma(pp(\bar{p})\rightarrow J/\psi\Upsilon)\times{\rm Br}(J/\psi\rightarrow \mu^+\mu^-){\rm Br}(\Upsilon\rightarrow \mu^+\mu^-)$ (in units of fb) of prompt $J/\psi$ and $\Upsilon(1S,2S,3S)$ simultaneous production at the Tevatron in the D0 fiducial region~\cite{Abazov:2015fbl} and at $\sqrt{s}=13$ TeV LHC in the LHCb acceptance $2<y_{J/\psi,\Upsilon}<4.5$, where we have also included feeddown contributions from higher-excited quarkonia decay.}
\label{tab:xsprompt} 
\end{normalsize}
\end{table*}

\medskip

\textit{Conclusions} --In this Letter, we have performed the first complete analysis of simultaneous production of prompt $\psi$ and $\Upsilon$ mesons, including all leading SPS contributions: namely, the complete $\mathcal{O}(\alpha_S^6)$ CS contribution, the $\mathcal{O}(\alpha_S^2\alpha^2)$ EW contribution, and the complete $\mathcal{O}(\alpha_S^4)$ COM contribution, as well as the feeddown contribution. For a long time, this process was considered to be a good probe of the COM. Our work shows that it is, in fact, most probably dominated by DPS contributions. The large variation of the $\mathcal{O}(\alpha_S^4)$ COM yield on the choice of LDMEs indicates such a process could be a good discriminator between different CO LDMEs on the market, provided that DPS contributions could be precisely subtracted, which seems not currently at reach. Yet, since some LDME sets yield to negative COM cross sections, the present process already disfavors these sets and is in any case extremely useful. Our computations also show that even though the SPS contributions are small, they are not completely negligible and one cannot systematically ignore them in the data. In particular, with our complete SPS yields, we obtained $\sigma_{\rm eff}\le 8.2$ mb at $68\%$ confidence level from the D0 measurement of prompt $J/\psi$ and $\Upsilon$ production at the Tevatron. Together with the lower-limit value $\sigma_{\rm eff}\ge 1.1$ mb extracted by the D0 Collaboration, it helps to test the (non)universality of $\sigma_{\rm eff}$ and to understand how poor the ``pocket formula" Eq.(\ref{eq:dps}) is in describing DPS in the future. Finally, we also present our predictions of prompt $J/\psi$ and $\Upsilon$ production at the LHCb.

\acknowledgments
We are grateful to Olga Gogota,  Jean-Philippe Lansberg, Peter Svoisky, and Zhenwei Yang for enlightening discussions. We especially thank Jean-Philippe Lansberg for reading the manuscript and for helping us to improve the write-up. H.-S.S is supported by
ERC grant No. 291377 \textit{LHCtheory: Theoretical predictions and analyses of LHC physics:
advancing the precision frontier}. Y.-J.Z is supported by the National Natural Science
Foundation of China (Grants No. 11375021), the New Century Excellent Talents in University (NCET) under grant NCET-13-0030,  the Major State Basic Research Development Program of China (No. 2015CB856701),  and the Fundamental Research Funds for the Central Universities.

\bibliography{huasheng}

\newpage 
\appendix 

\begin{widetext}

\section{Supplemental material}

\subsection{Predictions with D0 fiducial cuts}\label{D0}
The fiducial region of D0 measurements is described in Ref.~\cite{Abazov:2015fbl}. Both $J/\psi$ and $\Upsilon$ mesons are reconstructed via their decay $J/\psi(\Upsilon)\rightarrow \mu^+\mu^-$ with the muons are asked for at least have transverse momenta $p_T(\mu^\pm)>$ 2 GeV and pseudorapidity $|\eta(\mu^\pm)|<2.0$. Besides Fig.\ref{fig:dist}, our predictions for other distributions are displayed in Fig.\ref{fig:CompareDZero}.

\begin{figure}
\begin{center}
\subfloat[Invariant mass distribution]{\includegraphics[width=0.49\textwidth,draft=false]{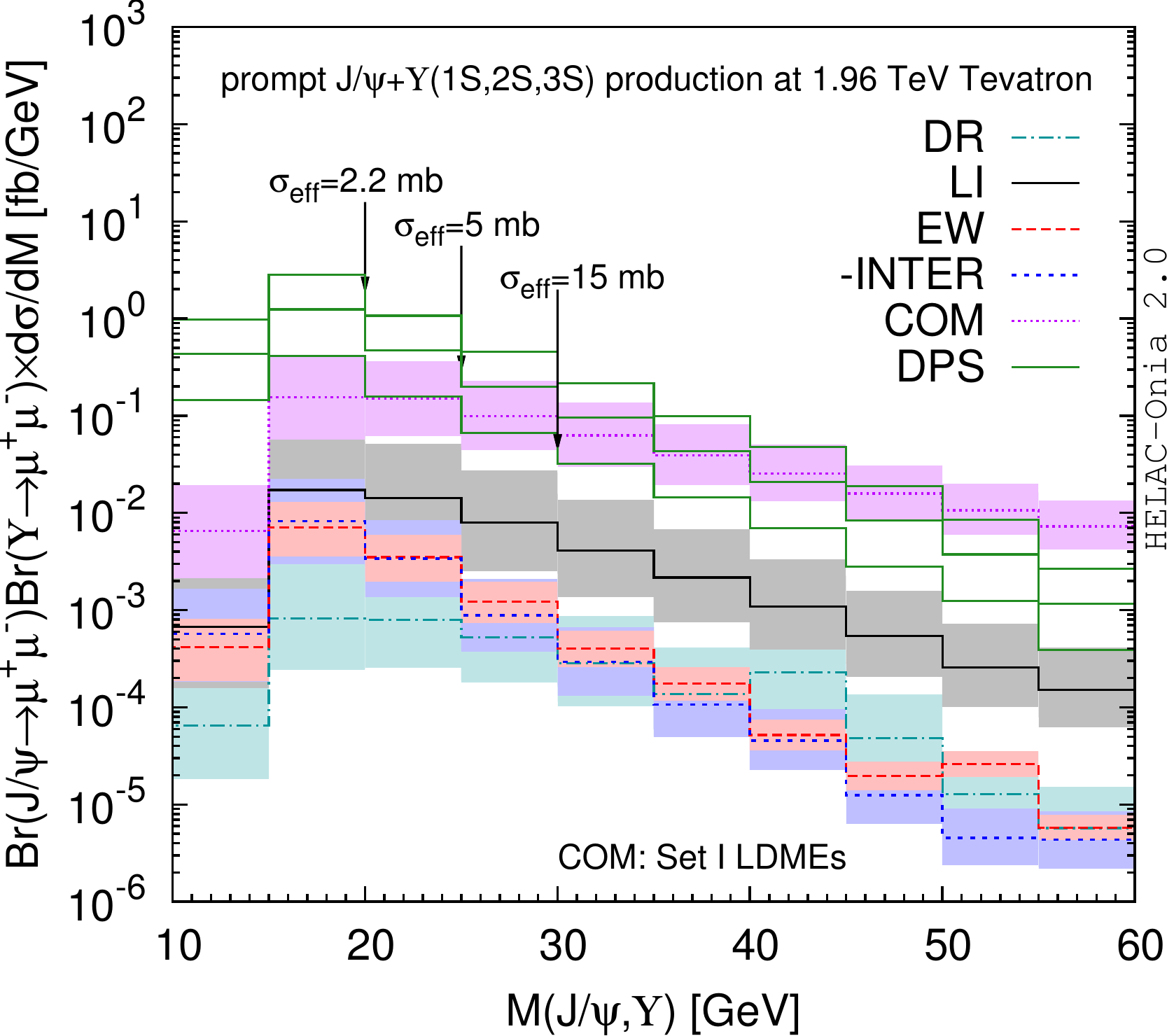}\label{fig:dsigDZeroa}}
\subfloat[Transverse momentum spectrum]{\includegraphics[width=0.49\textwidth,draft=false]{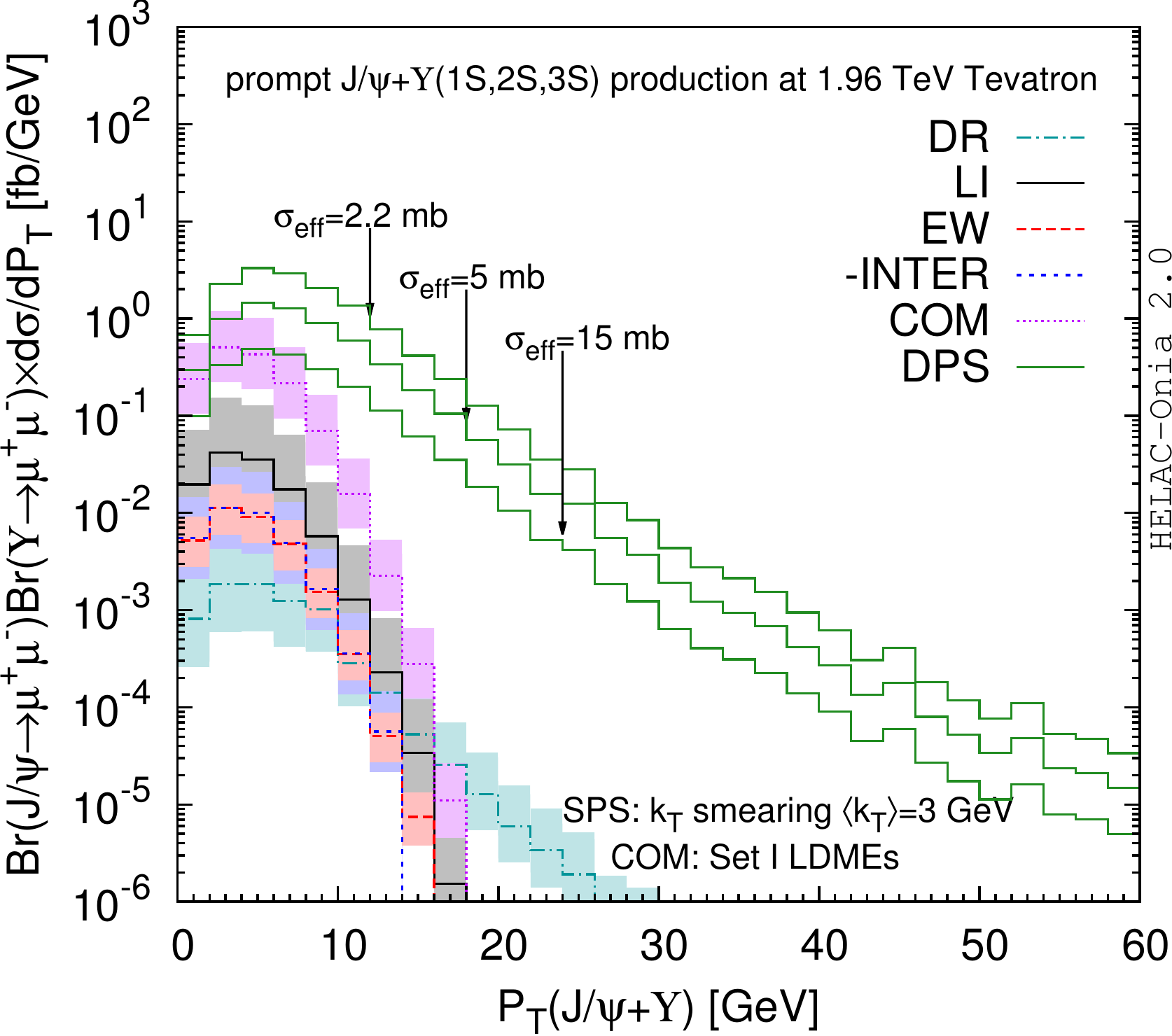}\label{fig:dsigDZerob}}\\
\subfloat[Transverse momentum of $\Upsilon$]{\includegraphics[width=0.49\textwidth,draft=false]{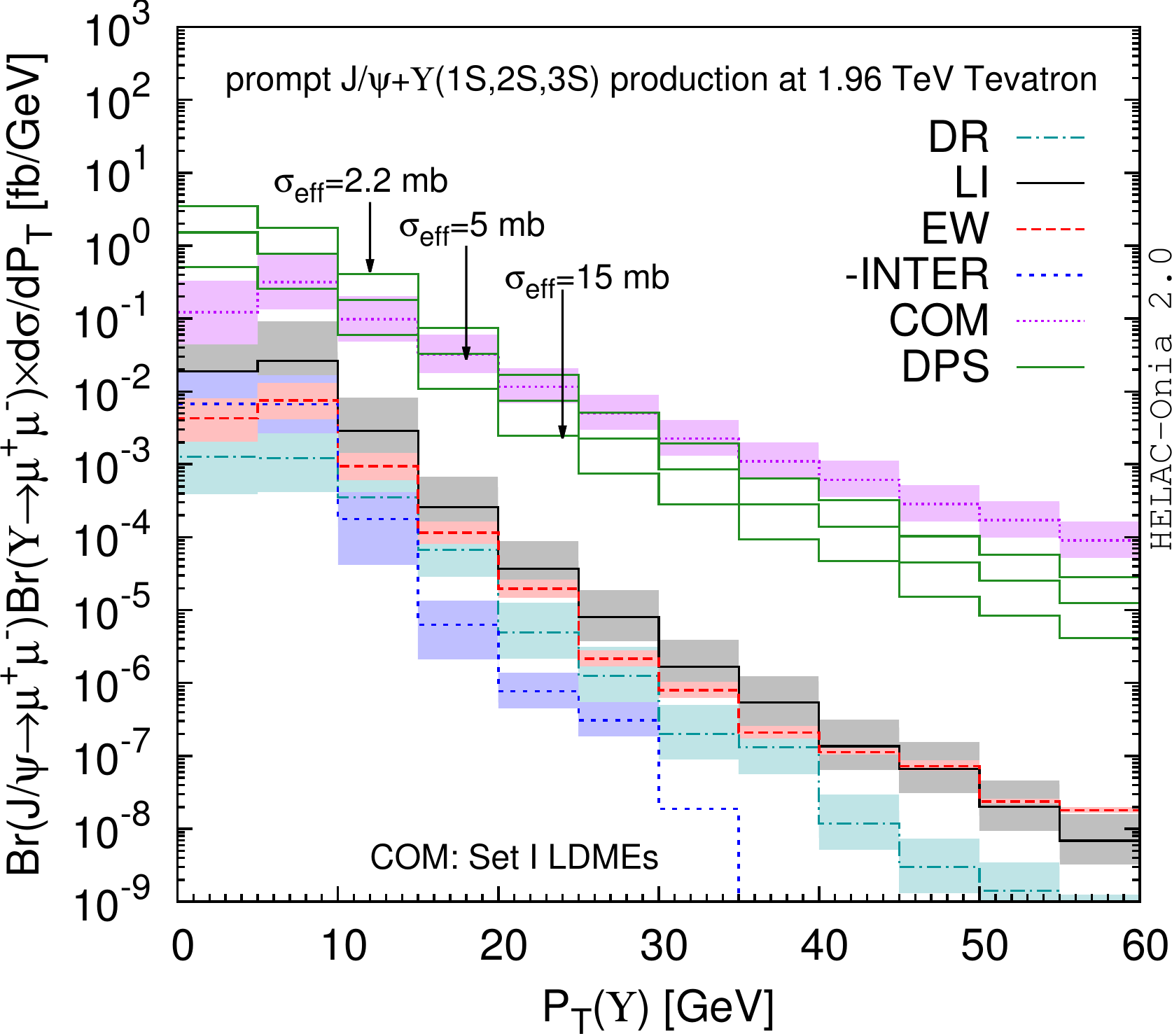}\label{fig:dsigDZeroc}}
\subfloat[Transverse momentum of $J/\psi$]{\includegraphics[width=0.49\textwidth,draft=false]{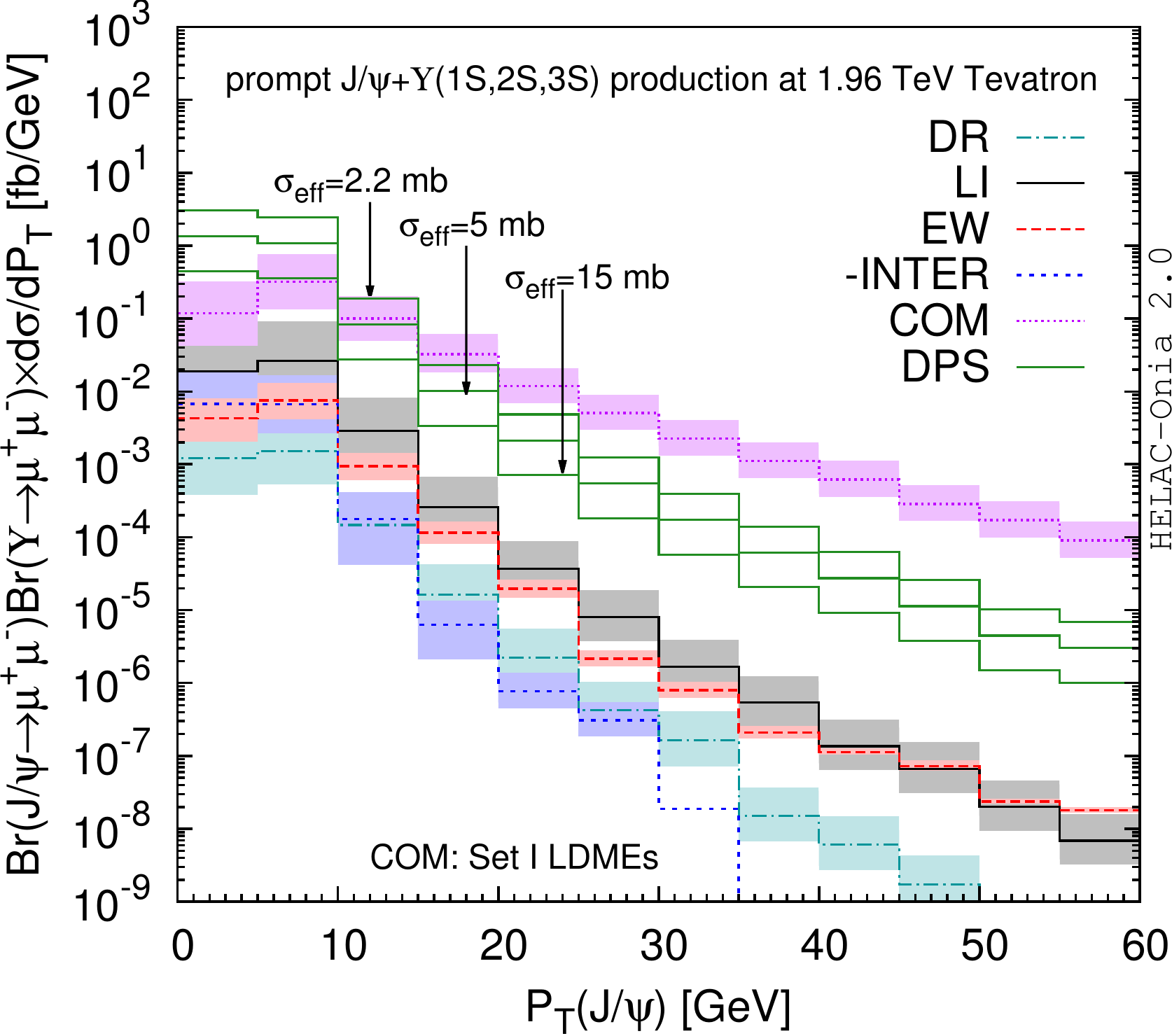}\label{fig:dsigDZerod}}
\caption{Predictions with D0 fiducial cuts:(a) invariant mass distribution; (b) transverse momentum spectrum; (c) transverse momentum distribution of $\Upsilon$; (d) transverse momentum distribution of $J/\psi$.
}
\label{fig:CompareDZero}
\end{center}\vspace*{-1cm}
\end{figure}

\subsection{Predictions with LHCb fiducial cuts}\label{LHCb}
After coordinating with the ongoing measurements performed by LHCb collaboration, we present our predictions in the new LHCb kinematic requirements. The results are displayed in Fig.\ref{fig:CompareLHCb}. In Table.\ref{tab:xsLHCbprompt}, we make a separation for $\Upsilon(1S,2S,3S)$ for the future comparison with experimental data.

\begin{table*}
\begin{normalsize}
\hspace{0cm}
\begin{tabular}{c|c|c|c|c|c|c|c|c}
\hline\rule{0pt}{3ex}
  Final states & \multicolumn{4}{c|}{CSM} & \multicolumn{4}{c}{COM}\\[1mm]
  & DR & LI & EW & INTER &
 Set I & Set II & Set III & Set IV \\[1mm] 
\hline\hline\rule{0pt}{3ex}
 $J/\psi+\Upsilon(1S)$ & $0.156^{+391\%}_{-79.7\%}$ &$3.69^{+436\%}_{-82.2\%}$
  & $1.04^{+75.5\%}_{-46.6\%}$ & $-1.97^{+162\%}_{-62.2\%}$ & $24.8^{+237\%}_{-72.8\%}$
  & $22.2^{+245\%}_{-73.9\%}$
  & $19.5^{+244\%}_{-73.8\%}$ 
 & $4.57^{+244\%}_{-73.9\%}$\\
\hline
\rule{0pt}{3ex}
 $J/\psi+\Upsilon(2S)$  & $0.0559^{+391\%}_{-79.7\%}$ & $1.33^{+436\%}_{-82.2\%}$
  & $0.375^{+135\%}_{-65.2\%}$ & $-0.709^{+262\%}_{-75.9\%}$ & $7.87^{+237\%}_{-72.8\%}$
  & $-2.94^{+256\%}_{-75.4\%}$
  & $5.42^{+241\%}_{-73.5\%}$ 
 & $1.36^{+242\%}_{-73.7\%}$\\
\hline
\rule{0pt}{3ex}
 $J/\psi+\Upsilon(3S)$  & $0.0434^{+391\%}_{-79.7\%}$ & $1.03^{+436\%}_{-82.2\%}$
  & $0.291^{+135\%}_{-65.2\%}$ & $-0.550^{+262\%}_{-75.9\%}$ & $6.07^{+238\%}_{-73.0\%}$
  & $1.98^{+241\%}_{-73.4\%}$
  & $3.15^{+242\%}_{-73.5\%}$ 
 & $0.634^{+240\%}_{-73.4\%}$\\
\hline
\end{tabular}
\caption{SPS cross sections $\sigma(pp\rightarrow J/\psi\Upsilon(nS))\times{\rm Br}(J/\psi\rightarrow \mu^+\mu^-){\rm Br}(\Upsilon(nS)\rightarrow \mu^+\mu^-),n=1,2,3$ (in unit of fb) of prompt $J/\psi$ and $\Upsilon(1S,2S,3S)$ simultaneous production at $\sqrt{s}=13$ TeV LHC in the LHCb acceptance $2<y_{J/\psi,\Upsilon}<4.5$, where we have also included feeddown contributions from higher-excited quarkonia decay.}
\label{tab:xsLHCbprompt} 
\end{normalsize}
\end{table*}

\begin{figure}
\begin{center}
\vspace{-1.2cm}
\subfloat[Azimuthal angle distribution]{\includegraphics[width=0.49\textwidth,draft=false]{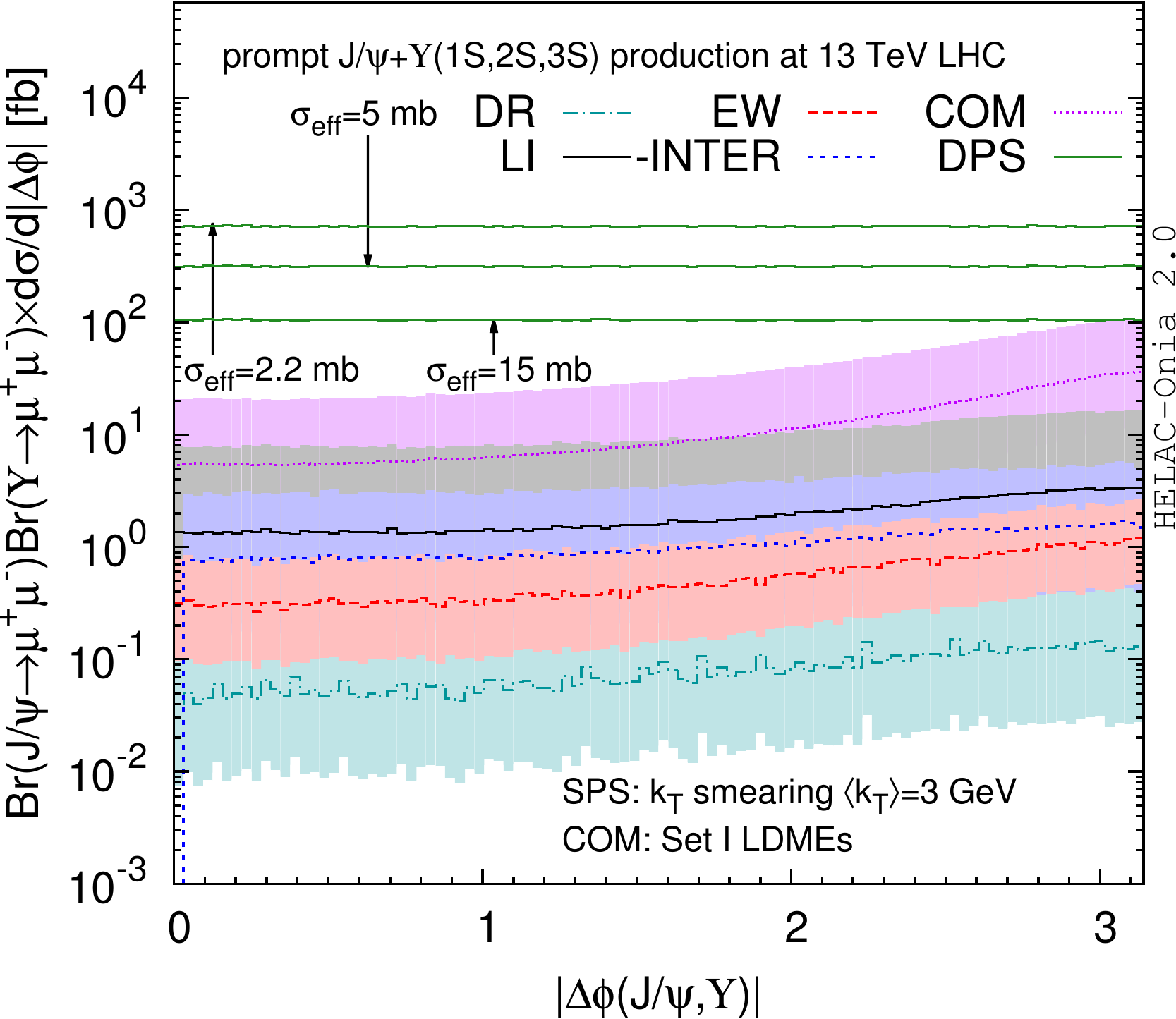}\label{fig:dsigLHCba}}
\subfloat[Rapidity difference distribution]{\includegraphics[width=0.49\textwidth,draft=false]{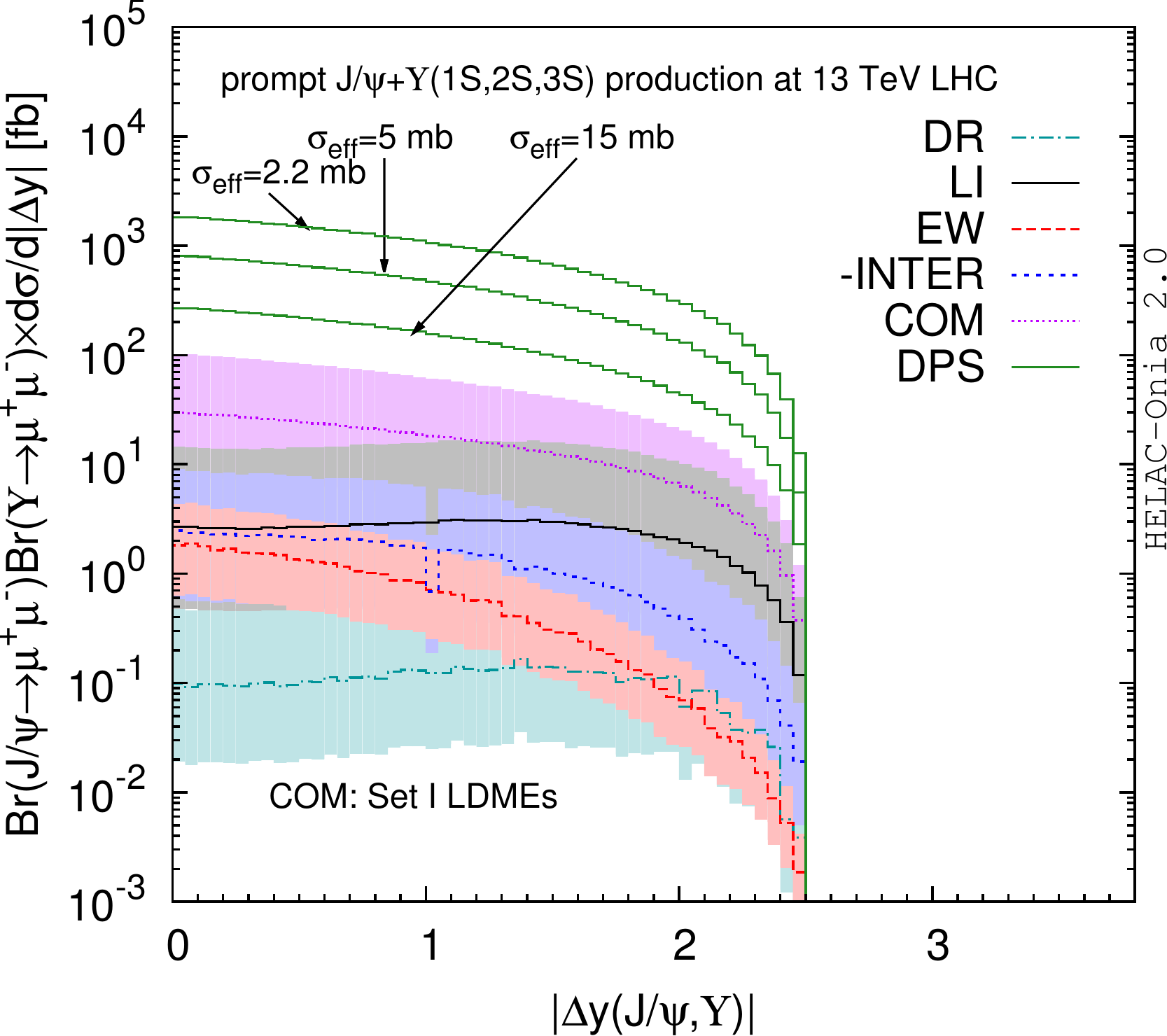}\label{fig:dsigLHCbb}}\\
\subfloat[Invariant mass distribution]{\includegraphics[width=0.49\textwidth,draft=false]{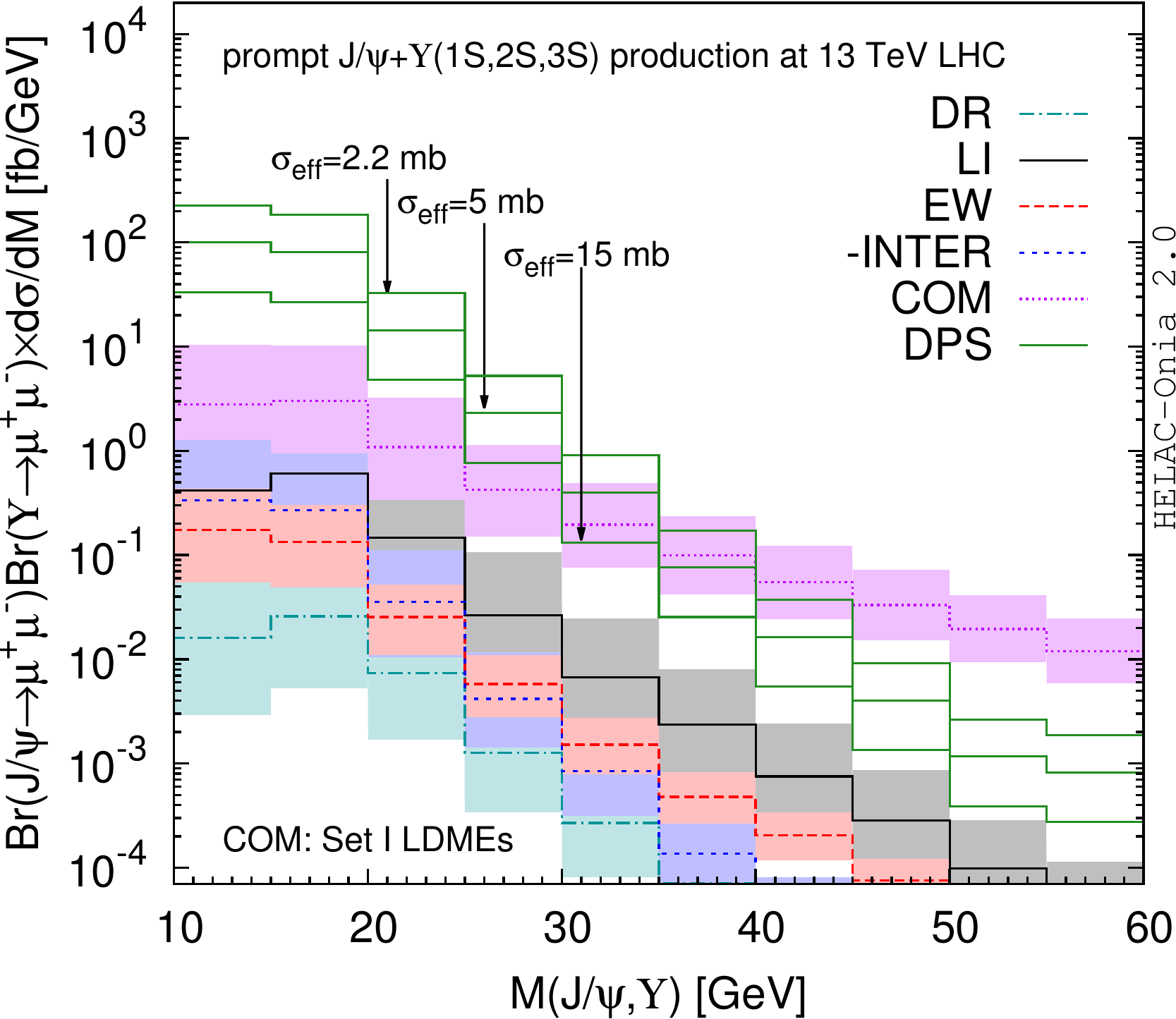}\label{fig:dsigLHCbc}}
\subfloat[Transverse momentum spectrum]{\includegraphics[width=0.49\textwidth,draft=false]{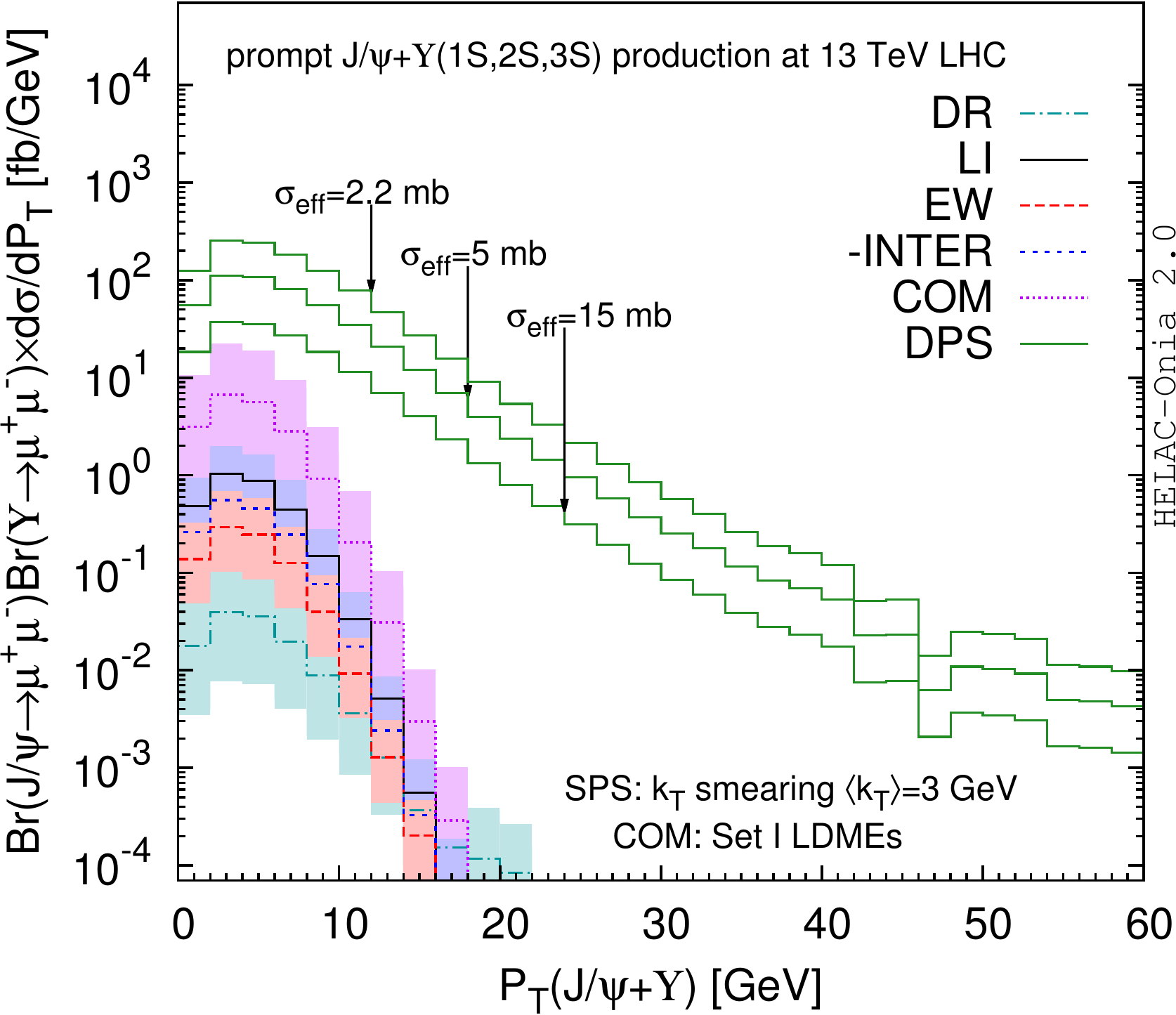}\label{fig:dsigLHCbd}}\\
\subfloat[Transverse momentum of $\Upsilon$]{\includegraphics[width=0.49\textwidth,draft=false]{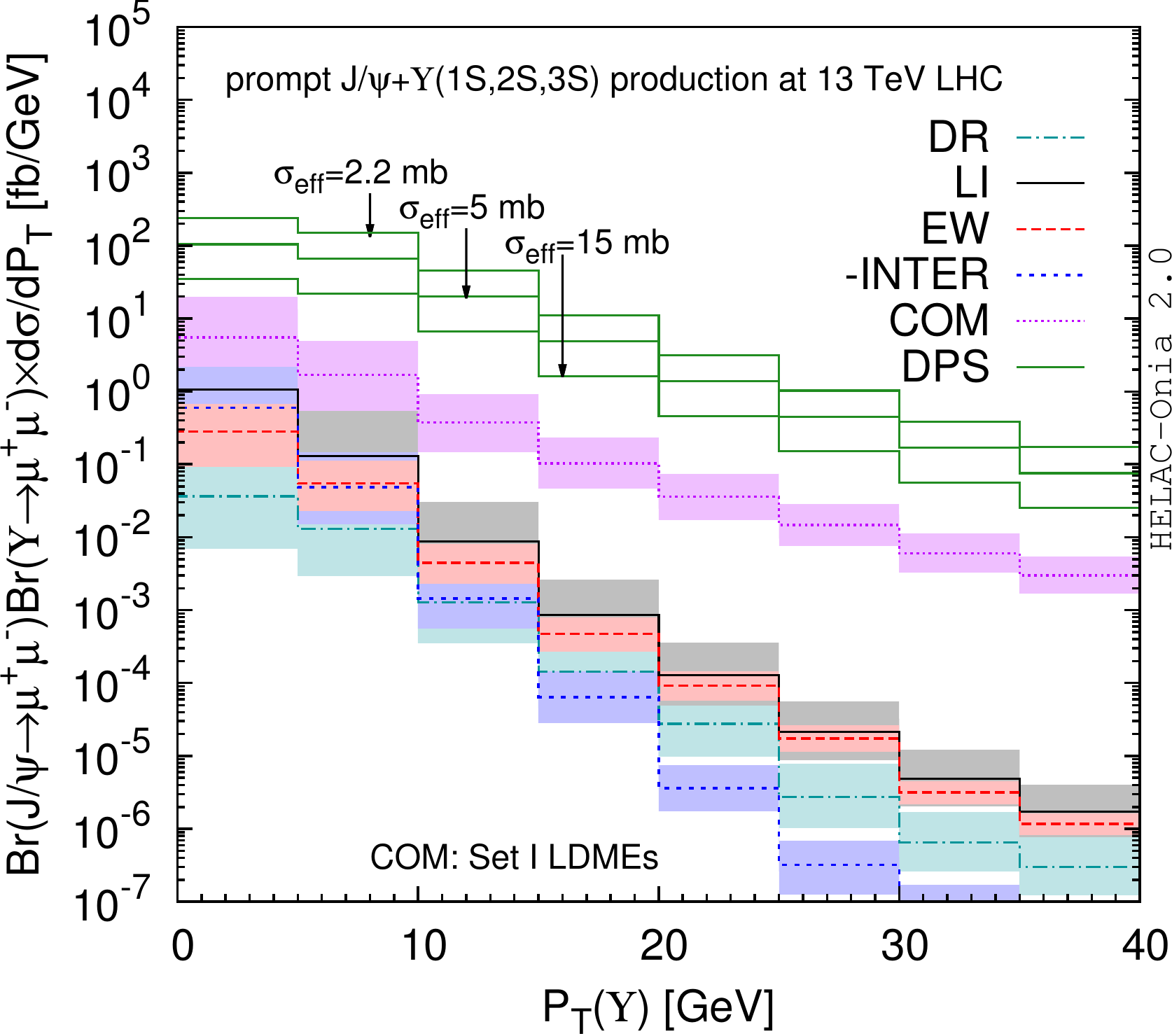}\label{fig:dsigLHCbe}}
\subfloat[Transverse momentum of $J/\psi$]{\includegraphics[width=0.49\textwidth,draft=false]{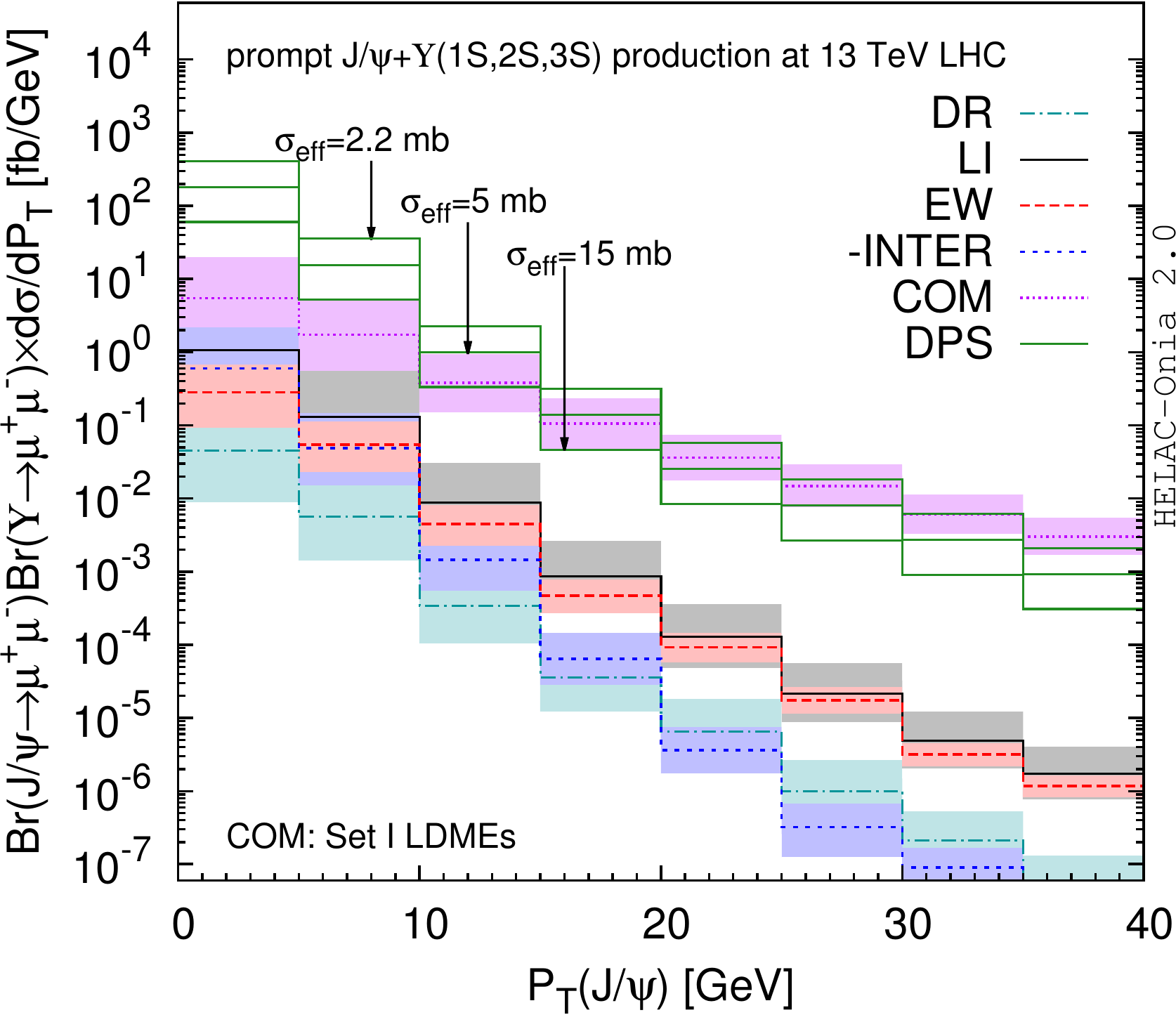}\label{fig:dsigLHCbf}}
\caption{Predictions with LHCb fiducial cuts:(a) azimuthal angle distribution; (b) rapidity difference distribution; (c) invariant mass distribution; (d) transverse momentum spectrum; (e) transverse momentum distribution of $\Upsilon$; (f) transverse momentum distribution of $J/\psi$.
}
\label{fig:CompareLHCb}
\end{center}\vspace*{-1cm}
\end{figure}

\end{widetext}

\end{document}